\documentclass[conference]{IEEEtran}
\usepackage{cite}
\usepackage{amsmath,amssymb,amsfonts}
\usepackage{graphicx}
\usepackage{textcomp}
\usepackage{xcolor}

\usepackage{mathtools}
\DeclarePairedDelimiter{\ceil}{\lceil}{\rceil}

\usepackage{algorithm} 
\usepackage[noend]{algpseudocode}

\usepackage[latin9]{inputenc}

\makeatletter \def\BState{\State\hskip-\ALG@thistlm} 

\def\ps@IEEEtitlepagestyle{%
  \def\@oddfoot{\mycopyrightnotice}%
  \def\@evenfoot{}%
}
\def\mycopyrightnotice{%
  {\footnotesize 978-1-5386-7568-7/18/\$31.00 \textcopyright 2018 IEEE\hfill}
  \gdef\mycopyrightnotice{}
}

\makeatother

\usepackage [autostyle, english = american]{csquotes}
\MakeOuterQuote{"}

\IEEEoverridecommandlockouts

\begin{document}

\title{Noise Flooding for Detecting Audio Adversarial Examples Against Automatic Speech Recognition\\
\thanks{This work is supported by the National Science Foundation under Grant No. 1659788.}
}

\author{\IEEEauthorblockN{Krishan Rajaratnam}
\IEEEauthorblockA{\textit{The College} \\
\textit{University of Chicago}\\
Chicago, USA \\
krajaratnam@uchicago.edu}
\and
\IEEEauthorblockN{Jugal Kalita}
\IEEEauthorblockA{\textit{Department of Computer Science} \\
\textit{University of Colorado}\\
Colorado Springs, USA \\
jkalita@uccs.edu}
}

\maketitle

\begin{abstract}
Neural models enjoy widespread use across a variety of tasks and have grown to become crucial components of many industrial systems. Despite their effectiveness and extensive popularity, they are not without their exploitable flaws. Initially applied to computer vision systems, the generation of adversarial examples is a process in which seemingly imperceptible perturbations are made to an image, with the purpose of inducing a deep learning based classifier to misclassify the image. Due to recent trends in speech processing, this has become a noticeable issue in speech recognition models. In late 2017, an attack was shown to be quite effective against the Speech Commands classification model. Limited-vocabulary speech classifiers, such as the Speech Commands model, are used quite frequently in a variety of applications, particularly in managing automated attendants in telephony contexts. As such, adversarial examples produced by this attack could have real-world consequences. While previous work in defending against these adversarial examples has investigated using audio preprocessing to reduce or distort adversarial noise, this work explores the idea of flooding particular frequency bands of an audio signal with random noise in order to detect adversarial examples. This technique of flooding, which does not require retraining or modifying the model, is inspired by work done in computer vision and builds on the idea that speech classifiers are relatively robust to natural noise. A combined defense incorporating 5 different frequency bands for flooding the signal with noise outperformed other existing defenses in the audio space, detecting adversarial examples with 91.8\% precision and 93.5\% recall.
\end{abstract}


\section{Introduction}

The growing use of deep learning models necessitates that those models be accurate, robust, and secure. However, these models are not without abusable defects. Initially applied to computer vision systems \cite{szegedy}, the generation of adversarial examples (loosely depicted in Fig. 1) is a process in which seemingly imperceptible changes are made to an image, with the purpose of inducing a deep learning based classifier to misclassify the image. The effectiveness of such attacks is quite high, often resulting in misclassification rates of above 90\% in image classifiers \cite{szegedy2}. Due to the exploitative nature of these attacks, it can \begin{figure}[htbp]
 \centerline{\includegraphics[width = 0.9\linewidth]{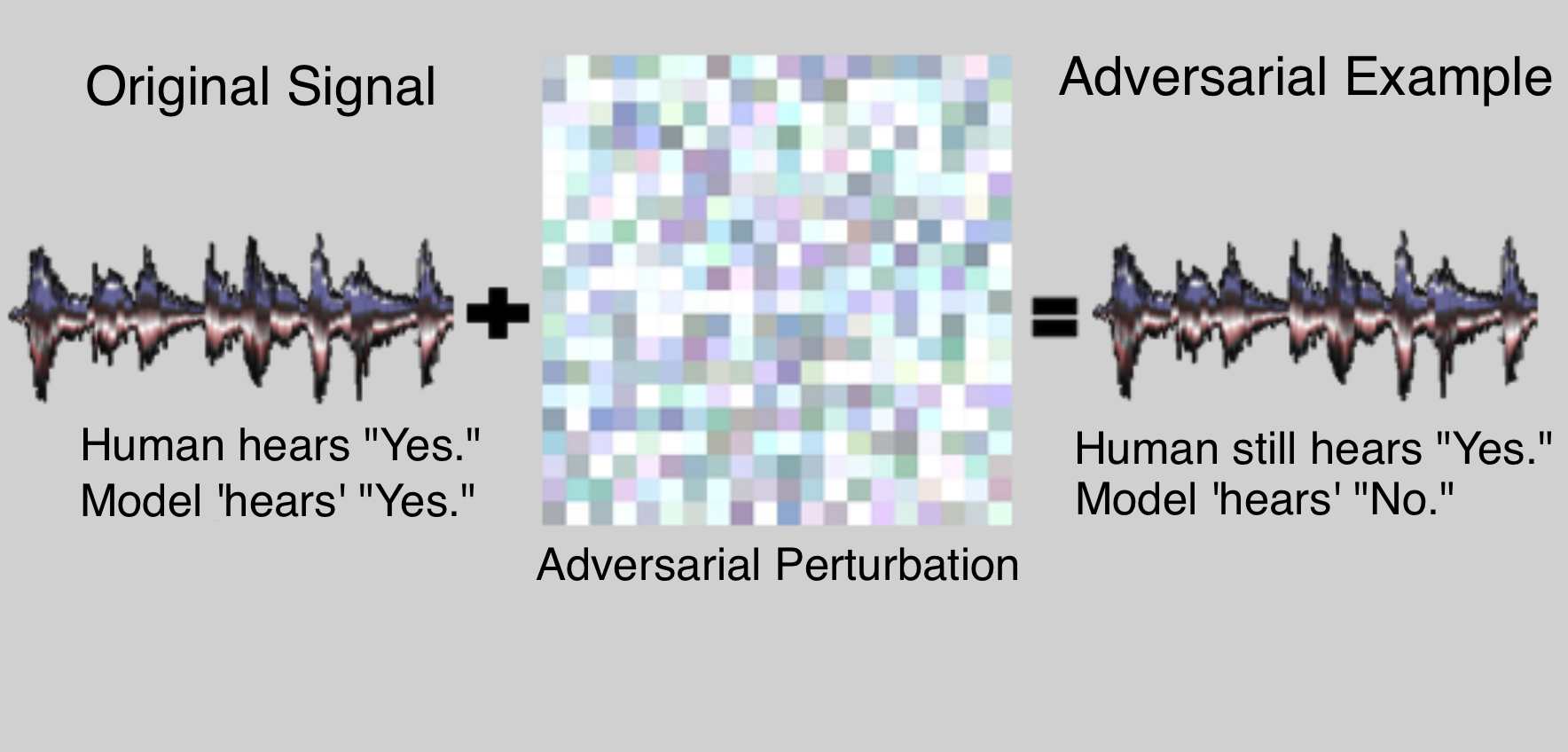}}
 \caption{A graphic depicting a targeted adversarial attack from "yes" (the source) to "no" (the target). A malicious attacker can add a small amount of adversarial perturbation to a signal such that it is classified by a model as some target class while a human still primarily hears the source class.}
 \label{fig}
\end{figure} be difficult to defend against adversarial examples while maintaining general accuracy.

The generation of adversarial examples is not just limited to image recognition. Although speech recognition traditionally relied heavily on hidden Markov models and various signal processing techniques, the gradual growth of computer hardware capabilities and available data has enabled end-to-end neural models to become more popular and even state of the art. As such, speech recognizers that rely heavily on deep learning models are susceptible to adversarial attacks. Recent work has been done on the generation of targeted adversarial examples against a convolutional neural network trained on the widely used Speech Commands dataset \cite{malzantot} and against Mozilla's implementation of the DeepSpeech end-to-end model \cite{carlini}, in both cases generating highly potent and effective adversarial examples that were able to achieve up to a 100\% misclassification rate. Due to this trend, the reliability of deep learning models for automatic speech recognition is compromised; there is an urgent need for adequate defense against adversarial examples.

\section{Related Work}

The attack against Speech Commands described by Alzantot et al. \cite{malzantot} is particularly relevant within the realm of telephony, as it could be adapted to fool limited-vocabulary speech classifiers used for automated attendants. This attack produces adversarial examples using a gradient-free genetic algorithm, allowing the attack to penetrate the non-differentiable layers of preprocessing typically used in automatic speech recognition.

\subsection{Audio Preprocessing Defenses} \label{lemmond}

As adversarial examples are generated by adding adversarial noise to a natural input, certain methods of preprocessing can serve to remove or distort the adversarial noise to mitigate the attack. 

Recent work in computer vision has shown that some preprocessing, such as JPEG and JPEG2000 image compression \cite{aydemir} or cropping and resizing \cite{grboult}, can be employed with a certain degree of success in defending against adversarial attacks. In a similar vein, preprocessing defenses have also been used for defending against adversarial attacks on speech recognition. Yang et al. \cite{yang2018towards} were able to achieve some success using local smoothing, down-sampling, and quantization for disrupting adversarial examples produced by the attack of Alzantot et al. While quantizing with $q = 256$, Yang et al. were able to achieve their best result of correctly recovering the original label of 63.8\% of the adversarial examples, with a low cost to general model accuracy. As quantization causes the amplitudes of sampled data to be rounded to the closest integer multiple of $q$, adversarial perturbations with small amplitudes can be disrupted.

Work has also been done in employing audio compression, band-pass filtering, audio panning, and speech coding to detect the examples of Alzantot et al. Rajaratnam et al. \cite{rsk} explored using these forms of preprocessing as a part of both isolated and ensemble methods for detecting adversarial examples. The discussed isolated preprocessing methods are quite simple; they merely check to see if the prediction yielded by the model is changed by applying preprocessing to the input. Despite this simplicity, Rajaratnam et al. achieved their best result of detecting adversarial examples with 93.5\% precision and 91.2\% recall using a "Learned Threshold Voting" (LTV) ensemble: a discrete voting ensemble composed of all of the isolated preprocessing methods that learns an optimal threshold for the number of votes needed to declare an audio sample as adversarial. They achieved a higher $F_1$ score for detecting adversarial examples using this voting ensemble when compared to more sophisticated techniques for combining the methods of preprocessing into an ensemble.

\subsection{Pixel Deflection}

While the aforementioned defenses focus on removing or distorting adversarial noise, one could also defend against an adversarial example by adding noise to the signal. Artificial neural network (ANN) classifiers are relatively robust to natural noise, whereas adversarial examples are less so. Prakash et al. \cite{prakash} used this observation and proposed a procedure for defending against adversarial images that involves corrupting localized regions of the image through the redistribution of pixel values. This procedure, which they refer to as "pixel deflection," was shown to be very effective for retrieving the true class of an adversarial attack. The strategy of defense proposed by Prakash et al. is more sophisticated than merely corrupting images by indiscriminately redistributing pixels; they target specific pixels of the image to deflect and also perform a subsequent wavelet-based denoising procedure for softening the corruption's impact on benign inputs. Regardless of the many aspects of the pixel deflection defense that seem to only be directly applicable to defenses within computer vision, the fundamental motivating idea behind this strategy---that ANN classifiers are robust to natural noise on benign inputs relative to adversarial inputs---is an observation that should also hold true for audio classification.

\section{Methods and Evaluation}  

Based off the observation of model robustness to natural noise, it should generally take less noise to change the model's prediction class of an adversarial example than it would to change that of a benign example. One could detect adversarial examples by observing how much noise needs to be added to the signal before the prediction that the model yields changes. Additionally, adversarial noise in audio is not localized to any particular frequency band, whereas much of the information associated with human speech is concentrated along the lower frequencies. As such, flooding particular frequency bands with random noise can be useful for detecting adversarial examples.

The aim of this research can be divided into two parts: testing the effectiveness of simple noise flooding (i.e. flooding the signal with randomly generated noise distributed along a particular band of frequency) for detecting audio adversarial examples, and combining multiple simple noise flooders that target different frequency bands together into an ensemble defense. The adversarial examples are produced using the gradient-free attack of Alzantot et al., against the same pre-trained Speech Commands model \cite{malzantot}. 

\subsection{Speech Commands Dataset and Model}

The Speech Commands dataset was first released in 2017 and contains 105,829 labeled utterances of 32 words from 2,618 speakers \cite{speechdataset}. This audio is stored in the Waveform Audio File Format (WAV) and was recorded  with a sample rate of 16 kHz. The Speech Commands model is a light-weight model based on a keyword spotting convolutional neural network (CNN) \cite{keywordcnn} that achieves a 90\% classification accuracy on this dataset. For the purposes of this research, a subset\footnote{The training and test datasets of adversarial and benign examples used in this research are available at http://github.com/LincLabUCCS/Noise-Flooding, along with the code used for implementing and testing the noise flooding defense.} of only 30,799 labeled utterances of 10 words are used, for consistency with previous work regarding the adversarial examples of Alzantot et al. From this subset, 20 adversarial examples are generated for each nontrivial source-target word pair, for a total of 1,800 examples. Each example is generated by implementing the attack with a maximum of 500 iterations through the genetic algorithm. Of these 1,800 generated examples, 128 are classified correctly (i.e. with the original source class) by the model. As such, only the remaining 1,672 examples (that are successful in fooling the model on some level) are used in this research.

\subsection{Simple Noise Flooding}  \label{ind}

This method for detecting adversarial examples involves calculating a score (that we term "flooding score") from an audio signal that represents how much random noise needs to "flood" the signal in order to change the model's prediction class of the audio signal. By calculating the flooding scores of the adversarial and benign examples in the training dataset, an ideal threshold score of maximum information gain can be found; test examples that have a flooding score less than the threshold are declared adversarial.

\subsubsection{Flooding Score Calculation}

Every audio signal can be represented as an array of $n$ audio samples along with a sample rate. A straightforward method of noising an audio signal with a noise limit $\epsilon$ is by generating an array of $n$ random integers between $-\epsilon$ and $\epsilon$ and adding this array to the original array of $n$ audio samples. The simple noise flooding defense noises audio in a similar manner, except $n$ random integers are passed through a band-pass filter before being added to the original array so that the added noise will be concentrated along a particular frequency band. The smallest $\epsilon$ found that induces a model prediction change between the noised signal and the original audio signal is used as a \textit{flooding score} for determining whether the original signal is an adversarial example. The procedure for calculating the flooding score of an audio signal is detailed in Algorithm 1.

\begin{algorithm}
\caption{Flooding Score Calculation Algorithm}\label{euclid}
\begin{algorithmic}[1]
\State \textbf{Input:} Audio signal $x$, model $m$, step size $s$, maximum noise level $\epsilon_{max}$, frequency band $b$
\State \textbf{Output:} Noise Flooding Score $\epsilon$
\State $n \gets$ number of samples in $x$
\State $pred_{orig} \gets$ classification of $x$ using $m$
\State $pred \gets pred_{orig}$
\State $\epsilon \gets 0$
\While {$pred = pred_{orig}$ and $\epsilon < \epsilon_{max}$}
\State $\epsilon \gets \epsilon + s$
\State $noise \gets$ $n$ uniform random integers taken from $[-\epsilon, \epsilon]$
\State apply band pass filter on $noise$ using $b$ 
\State $pred \gets$ classification of $x + noise$ using $m$
\EndWhile
\State \Return $\epsilon$
\end{algorithmic}
\end{algorithm}

This procedure will make no more than $\ceil{\epsilon_{max}/s}$ calls to the model when calculating the flooding score of an audio signal. As such, there is an inherent trade-off that comes with the choice of the step size parameter $s$; a large step size would generally cause the algorithm to terminate quickly with a less precise score, whereas a small step size would result in a more precise score but at a higher computational cost. In this research, a step size of $50$ was used, though in practice this parameter could be tuned to suit particular scenarios.  A similar trade-off is implicit with the choice of the $\epsilon_{max}$ parameter. 

\subsubsection{Frequency Bands for Testing}

The simple noise flooding procedure can be tested using various bands of frequency for concentrating noise. Considering that the sample rate of files within the Speech Commands dataset is 16 kHz, the Nyquist frequency \cite{nyquist} of this system is 8000 Hz. Considering that the 0-8000 Hz frequency range can be divided into 4 bands of equal width, we are left with the following 5 variations of simple noise flooding for testing:

\begin{itemize}
\item Unfiltered Noise Flooding,
\item 0-2000 Hz Noise Flooding,
\item 2000-4000 Hz Noise Flooding,
\item 4000-6000 Hz Noise Flooding, and  
\item 6000-8000 Hz Noise Flooding.
\end{itemize}

It is worth noting that for unfiltered noise flooding, the noise array is not passed through any band pass filter. As such, the frequency band parameter $b$ (and, along with it, line 10 of Algorithm 1) is unused for calculating an unfiltered noise flooding score.

\subsection{Ensemble Defense}

While the above variations of the simple noise flooding defense may be somewhat effective for detecting adversarial examples in isolation, a more robust defense would be to combine the variations into an ensemble. As flooding scores calculated for each band may contain unique information that could be valuable for detecting adversarial examples, a defense that incorporates different varieties of flooding scores should be more effective. The flooding scores can be combined in a variety of configurations.

\subsubsection{Majority Voting}

A somewhat naive, yet direct, approach for combining the simple noise flooding variations is to use a discrete voting ensemble: for every audio signal passed, perform each of the 5 variations of simple noise flooding and tally up the adversarial "votes" each of the methods yield. If there are 3 or more (i.e. a majority) adversarial votes, the signal is declared adversarial.

\subsubsection{Learned Threshold Voting}

This ensemble technique is identical to the homonymous method described in \cite{rsk}. Although the majority voting technique requires 3 adversarial votes (i.e. a majority) for an adversarial declaration, this voting threshold is arbitrary. The learned threshold voting technique assesses the performance of voting ensembles using all possible voting thresholds on a training dataset, and chooses the threshold that yielded the best performance. For quantifying performance, $F_1$ scores are used, though one could adjust this $F$-measure to accommodate one's outlook on the relative importances of recall and precision.

\subsubsection{Tree-Based Classification Algorithms}

The previous ensemble techniques do not discriminate between voters in the ensemble; every vote is considered equal. Considering that human speech information is not distributed evenly among the frequency bands used in the noise flooding ensemble (most human speech information would be distributed along the 0-2000 Hz band), it may be somewhat callow to treat each member of the ensemble equally.

Decision tree-based classification algorithms generally perform well in classifying vectors of features into discrete classes. To avoid discarding information, one could calculate the simple flooding score yielded by each member of the ensemble and concatenate these scores into a 5-dimensional flooding score vector and train a tree-based classification algorithm to detect adversarial examples from its flooding score vector. In this work, 3 tree-based classification algorithms will be used, due to their high performance on a variety of discrete classification tasks:

\begin{itemize}
\item Adaptive Boosting (AdaBoost) \cite{adaboost},
\item Random Forest Classification \cite{randomforest}, and
\item Extreme Gradient Boosting (XGBoost) \cite{xgboost}.
\end{itemize}




\subsection{Evaluation}

All of the previously mentioned detection methods are evaluated based off their precisions and recalls in detecting adversarial examples from a test set of 856 adversarial examples and 900 benign examples---the remaining 816 adversarial examples and an additional 900 benign examples are used to calculate flooding scores for training.

When applying defenses against adversarial examples, an implied tradeoff between the general usability and security of the model seems to arise. From a security standpoint, it is extremely important to have a high recall in detecting adversarial examples, whereas for the sake of general usability, there should be a high precision when declaring a potentially benign input as adversarial. This research takes the stance that both general usability and security are equally important. As such, $F_1$ scores are used when evaluating the defenses in order to equally balance precision and recall.

\section{Results}

The precisions, recalls, and $F_1$ scores are evaluated for each of the simple noise flooding defenses in addition to the two best isolated preprocessing defenses from \cite{rsk} (i.e. the two isolated defenses with the highest $F_1$ scores) and are shown in Table \ref{tab0}. \begin{table}[htbp]
\caption{Performance of Simple Noise Flooding Defenses}
\begin{center}
\begin{tabular}{ | p{4.0cm} | p{1cm} | p{1cm} | p{0.6cm} |}
    \hline
    Detection Method & Precision & Recall & $F_1$ Score \\ \hline
Unfiltered Noise Flooding & 89.8\% & 93.1\% & \textbf{0.914}\\ \hline
0-2000 Hz Noise Flooding & 88.3\% & \textbf{94.5\%} & 0.913 \\ \hline
2000-4000 Hz Noise Flooding & 88.3\% & 92.5\% & 0.905 \\ \hline
4000-6000 Hz Noise Flooding  & 86.3\% & 92.5\% & 0.893\\ \hline
6000-8000 Hz Noise Flooding & 82.0\% & 91.5\% & 0.865\\ \hline
Isolated Speex Compression$^{{\mathrm{a}}}$& 93.7\% & 88.5\% &  0.910\\ \hline
Isolated Panning \& Lengthening$^{{\mathrm{a}}}$ & \textbf{95.8\%} & 82.4\%  & 0.886\\ 
    \hline 
    \multicolumn{4}{l}{$^{\mathrm{a}}$Taken from Rajaratnam et al.} \\ 
    \end{tabular} 
\label{tab0}
\end{center}
\end{table} 
\pagebreak
\begin{figure}[htbp]
 \centerline{\includegraphics[width = 0.9\linewidth]{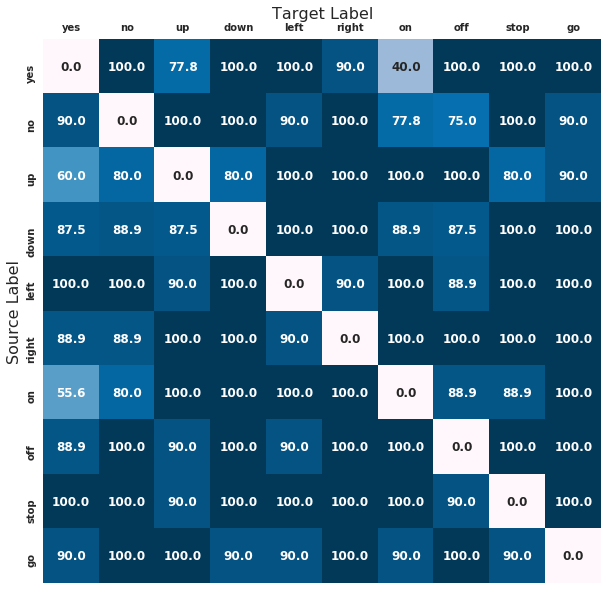}}
 \caption{A heat map depicting recall values (as percentages) for detecting audio adversarial examples using the noise flooding extreme gradient boosting ensemble. The diagonal of zeroes correspond to trivial source-target pairs for which there were no adversarial examples generated.}
 \label{fig}
\end{figure} 
From the results, one can see that the simple noise flooding defenses are all able to achieve higher recalls than the isolated preprocessing detection methods proposed by Rajaratnam et al. in \cite{rsk}. 

Additionally, the simple noise flooding methods that target lower frequency bands performed better than those that targeted higher frequency bands. This frequency-based disparity in performance follows reasonably from the fact that human speech 
information is concentrated in the lower frequencies. While the unfiltered noise flooding method achieved the highest $F_1$ score, the 0-2000 Hz Noise Flooding defense achieved a higher recall in detecting adversarial examples.


The results of the ensemble noise flooding defenses in addition to the two best ensemble preprocessing defenses from \cite{rsk} are summarized in Table \ref{tab1}. Most of the ensemble techniques achieve higher $F_1$ scores than any of the individual simple flooding defenses. Understandably, the somewhat naive noise flooding majority voting ensemble yielded the lowest $F_1$ score of all the ensemble techniques. The noise flooding learned threshold voting ensemble \smallskip \begin{table}[htbp]
\caption{Performance of Ensemble Defenses}
\begin{center}
\begin{tabular}{ | p{4.0cm} | p{1cm} | p{1cm} | p{0.6cm} |}
    \hline
    Detection Method & Precision & Recall & $F_1$ Score \\ \hline
Noise Flooding Majority Voting & 88.0\% & 93.6\% & 0.907 \\ \hline
Noise Flooding LTV$^{{\mathrm{a}}}$ & 90.8\% & 92.2\% & 0.915 \\ \hline
Noise Flooding Random Forest & 90.9\% & 93.1\% & 0.920 \\ \hline
Noise Flooding AdaBoost  & 90.3\% & \textbf{94.2\%} & 0.922\\ \hline
Noise Flooding XGBoost & 91.8\% & 93.5\% & \textbf{0.926}\\ \hline
Preprocessing Majority Voting$^{{\mathrm{b}}}$& \textbf{96.1\%} & 88.1\% &  0.919\\ \hline
Preprocessing LTV$^{{\mathrm{a}}}$ $^{{\mathrm{b}}}$& 93.5\% & 91.2\% &  0.924\\ 
    \hline 
    \multicolumn{4}{l}{$^{\mathrm{a}}$LTV is short for the discrete Learned Threshold Voting ensemble.} \\ 
    \multicolumn{4}{l}{$^{\mathrm{b}}$Taken from Rajaratnam et al.} \\ 
    \end{tabular} 
\label{tab1}
\end{center}
\end{table} improves from the majority voting ensemble by learning a new voting threshold  of 4 (as opposed to 3, which is used in the majority voting ensemble). This higher threshold results in a lower recall in detecting adversarial examples, but results in a markedly higher precision in order to achieve an overall higher $F_1$ score.


As expected, the tree-based classification algorithms were the most effective for combining the simple noise flooding methods together, as they were able to learn an optimal method for discriminating between the members of the ensemble
 while the voting ensembles implicitly treated each voter equally. The adaptive boosting ensemble achieved a higher recall than any of the other ensemble noise flooding defenses, whereas the extreme gradient boosting ensemble achieved the highest $F_1$ score of any detection method. The recall measurements for detecting adversarial examples using the noise flooding extreme gradient boosting ensemble are detailed in Fig. 2.

\section{Conclusion and Future Work}

Although the results suggest that an ensemble noise flooding defense is effective in defending against adversarial examples produced by the unmodified algorithm of Alzantot et al., it does not necessarily show that this defense is secure against more complex attacks. While an ensemble defense may provide marginal security over the simple noise flooding methods in isolation, recent work has shown adaptive attacks on image classifiers are able to bypass ensembles of weak defenses \cite{carlini2}; this work could be applied to attack speech recognition models. Future work can be done to adapt noise flooding into a stronger defense that can withstand these types of adaptive adversarial examples, or at least cause the attacks to become more perceptible.

Additionally, this paper only discusses flooding signals with random noise that is effectively sampled from a uniform distribution. Future work can be done in exploring other techniques for producing the noise, perhaps by sampling from a more sophisticated probability distribution or deflecting individual samples. 

While the noise flooding techniques were able to yield high recalls and overall $F_1$ scores for detecting adversarial examples, many of the preprocessing-based defenses described in \cite{rsk} yielded higher precisions. This suggests that a defense that combines aspects of those defenses with noise flooding may be quite effective in detecting adversarial examples.

Prakash et al. \cite{prakash} softened the effect that their pixel deflection defense had on benign inputs by applying a denoising technique after locally corrupting the images. Perhaps a denoising technique could be applied after noise flooding to produce a more sophisticated defense that would yield a higher precision. 

Future work could also be done in adapting noise flooding into a defense that can restore the original label of adversarial examples, rather than simply detecting adversarial examples.

This paper proposed the idea of noise flooding for defending against audio adversarial examples and showed that fairly simple flooding defenses are quite effective in detecting the single-word targeted adversarial examples of Alzantot et al. This paper also showed that simple noise flooding defenses can be effectively combined together into an ensemble for a stronger defense. While these defenses may not be extremely secure against more adaptive attacks, this research aimed ultimately to further discussion of defenses against adversarial examples within the audio domain: a field in desperate need of more literature. 

\section*{Acknowledgments}
We are thankful to the reviewers for helpful criticism, and the UCCS LINC and VAST labs for general support. We also acknowledge the assistance of Viji Rajaratnam in creating Fig. 1.

\bibliographystyle{IEEEtran}
\bibliography{rak_adv_voip}

\end{document}